\documentclass[3p, preprint, times]{elsarticle}
\graphicspath{ {./figures/} }
\usepackage{natbib}
\usepackage{hyperref}
\usepackage{float}
\usepackage{amsmath}
\usepackage{amssymb}
\usepackage{verbatim} 
\usepackage{apalike}
\usepackage{mathrsfs,amsmath}
\restylefloat{figure}
\restylefloat{table}
\hypersetup{colorlinks}
\usepackage{color}
\usepackage{array}
\usepackage{algpseudocode}
\usepackage{algorithm}

\usepackage{booktabs}
\usepackage{multirow}

\biboptions{authoryear}

\begin{document}
\begin{frontmatter}

\journal{Neurocomputing}

\title{STAR: \\ A Session-Based Time-Aware Recommender System}

\author[label1]{Reza Yeganegi}
\ead{yeganegi.reza@ut.ac.ir}

\author[label1]{Saman Haratizadeh\corref{cor1}}
\ead{haratizadeh@ut.ac.ir}

\cortext[cor1]{Corresponding author}
\address[label1]{Faculty of New Sciences and Technologies, University of Tehran, Tehran, Iran}

\begin{abstract}

Session-Based Recommenders (SBRs) aim to predict users' next preferences regard to their previous interactions in sessions while there is no historical information about them. Modern SBRs utilize deep neural networks to map users' current interest(s) during an ongoing session to a  latent space so that their next preference can be predicted. Although state-of-art SBR models achieve satisfactory results, most focus on studying the sequence of events inside sessions while ignoring temporal details of those events. In this paper, we examine the potential of session temporal information in enhancing the performance of SBRs, conceivably by reflecting the momentary interests of anonymous users or their mindset shifts during sessions. We propose the STAR framework, which utilizes the time intervals between events within sessions to construct more informative representations for items and sessions. Our mechanism revises session representation by embedding time intervals without employing discretization. Empirical results on Yoochoose and Diginetica datasets show that the suggested method outperforms the state-of-the-art baseline models in Recall and MRR criteria.

\end{abstract}


\end{frontmatter}

\label{sec:Introduction}

A Session-based recommender system aims to predict the users' next item based on their previous interacted items in sessions. Such a system has two characteristics that distinguish it from other recommender systems: 1) The lack of users' identity information and 2) the significance of short-term preferences \cite{wang2019survey}.

Since the only information source is the interaction data in an ongoing session, SBRs suffer from users' identity unavailability. Thus, user-item matrix-analysis models cannot be used directly in this class of recommender systems. Furthermore, compared to a wide range of recommendation models that employ all historical users' intentions, SBR models recommend items by focusing on users' interactions in the current session. 

In recent years deep learning approaches have been one of the most potent methods. These models usually concentrate on encoding a session into a vector defined as a session representation resulting from the recommendation errors. Seminal studies on deep learning models \cite{zhang2014sequential,hidasi2016session} tried to design SBRs with different types of deep neural networks to outperform the conventional models. These models are extended by using techniques like data augmentation \cite{tan2016improved}, attention mechanism \cite{li2017neural,liu2018stamp}, combining the conventional models with deep learning models \cite{jannach2017recurrent}, and feature engineering \cite{wu2019session}.

\begin{figure}[h]
    \begin{center}
        \includegraphics[width=0.7\textwidth]{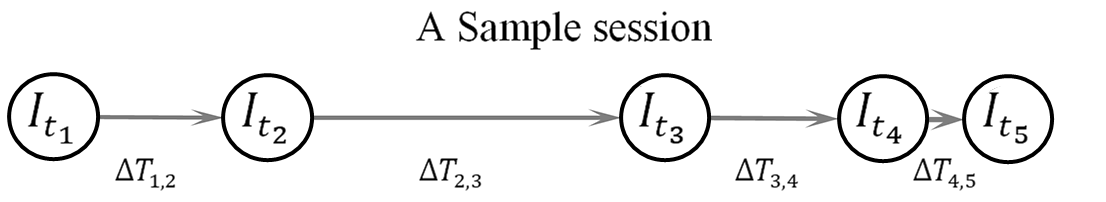}\\
        \caption{Time Intervals Between Interactions}
        \label{fig:Time_Interval}
    \end{center}
\end{figure}

One aspect of the session-based data that has been investigated rarely is the role of time intervals between user-item interactions in reflecting users' mindset during sessions. Intuitively, those time intervals may be interpreted in several ways: the importance of an item, distraction of a user, a change in a user mindset, an unintentional click, etc. As a case in point, please see the session presented in Fig.~\ref{fig:Time_Interval}. The time intervals between items $t_2$ and $t_3$ are considerable; therefore, the user has probably spent much time on item $t_2$. This observation can indicate his/her interest, which increases the chance of revisiting this item in the session. On the other hand, a relatively short interaction time (e.g., $\Delta T_{4,5}$) may show a lack of interest.

However, it is clear that, usually, there is no simple explanation for such an observation. For example, a substantial time interval between interactions may indicate a temporary user distraction or a technical problem while visiting the web page. Therefore, we probably need comprehensive models to combine the time interval data and other available information in a session to extract the user's interests.

In this paper, we present the STAR framework (Session-based and Time-Aware Recommender system) that employs the sequence of items observed in a session, as well as the time intervals between them, to construct a vector representation for a session, which accurately reflects the interest of the user at the moment and then uses this representation to predict his/her subsequent interest. We also utilize a technique in which there is no need for any discretization method of time interval values. In this system, time intervals are one of the main components that directly influence outputs. 

The main contributions of this work are summarized as follows:

\begin{enumerate}
    \item We introduce an approach to using time intervals between user-item interactions in a session to learn better behavioral patterns and make more accurate recommendations.

	\item We propose an approach to embed time intervals between items and utilize it to adjust each item's effect upon the aggregated session representation. 
	
	\item We suggest a method that uses time interval information to construct temporally sub-sessions in order to embed items based on their co-occurrence.
	
\end{enumerate}

The rest of this paper is organized as follows: In Section (\ref{sec:background}), we will review the existing session-based recommender systems. Then in section (\ref{sec:proposedModel}), the STAR framework and the training steps will be described. The experimental results will be reported, and our model will be compared to the state-of-the-art baselines in section (\ref{sec:results}). Finally, in section (\ref{sec:conclusion}), we will conclude the paper and suggest future works.

\section{Related Works}
\label{sec:background}

\textit{Conventional approaches:} Conventional session-based recommender systems usually analyze a part of sessions to infer the next step recommendations, namely K-Nearest Neighbor (KNN) based approaches, which are simple but effective \cite{ludewig2018evaluation}. Their primary idea is to find the similarity between the current session and previous ones in train data in various ways. For example, some of them seek the similarity between the last items of the sessions (Item-based KNN) \cite{hidasi2016session} or the entire events inside the sessions (Session-based KNN) \cite{jannach2017recurrent}.

Markov chain models are other types of conventional SBRs that utilize the Markov Decision Process (MDP). They consider sessions as Markov chains and endeavor to predict the next item based on transition probability between items. Whether the transition probability is calculated based on direct observations or latent spaces, Markov models can be divided into basic Markov chain-based approaches and latent Markov embedding-based approaches. \cite{zhang2007efficient} combined first and second-order Markov transitions to assemble a more accurate recommender model in the Web page domain. \cite{chen2012playlist} presented Logistic Markov Embedding (LME) to represent songs as one (or multiple) point(s) in Euclidean space so that Euclidean distance between songs reflects the transition probabilities between songs.

A well-known class of recommender systems includes methods that build latent representations by user and item connections. These methods aim to find patterns between users and items but are powerless for SBRs with anonymous users. However, with a few modifications, they can be applied to SBRs. \cite{rendle2010factorizing} proposed Factorized Personalized Markov Chains (FPMC) for the next basket recommendation. By limiting the basket size to one and considering the current session as the basket history, FPMC can be directly applied to SBRs \cite{ludewig2018evaluation}. Factored Item Similarity Model (FISM) is another model based on item-item factorization \cite{kabbur2013fism}. \cite{ludewig2018evaluation} used FISM with a few modifications from the movie recommendations domain to SBRs. Factorized Sequential Prediction with Item SImilarity modeL (FOSSIL) \cite{he2016fusing} combined FISM and FPMC and introduced another Latent representation-based approach.

\textit{Deep Learning approaches:} In recent years applying neural networks, especially RNNs, to improve the performance of SBRs has gained significant attention. The main idea of these methods is to learn different levels of behavioral patterns. \cite{hidasi2016session} offered GRU4Rec, which operates RNNs and suggests the session parallel mini-batch method with negative sampling to feed sessions online into a GRU. GRU4Rec achieved remarkable results as opposed to Conventional SBRs. GRU4Rec generates a representation for each session which is the last hidden state of the GRU. Finally, this representation lists a set of items as a recommendation. They also introduced pairwise ranking loss functions, leading to better results than cross-entropy loss. Subsequent attempts have been made to increase the accuracy of GRU4Rec. \cite{tan2016improved} improved GRU4Rec using data augmentation techniques. \cite{jannach2017recurrent} presented a heuristics-based nearest neighbor (KNN) scheme for sessions by combining RNN and KNN. The power of modeling the short-term preferences of SBRs can be used to improve the accuracy of the other types of recommenders when the users' IDs are available. \cite{zhao2019leveraging} utilized RNN to model short-term preferences and MF for long-term interests (that change very slowly) with the Generative Adversarial Network (GAN) in the movie recommendation domain.

NARM \cite{li2017neural} and STAMP \cite{liu2018stamp} were designed to model short-term and long-term users' preferences simultaneously in sessions. They argue that models which have been proposed merely based on the last hidden of RNN do not distinguish between items priority in sessions. However, a user may change his/her preferences during a session (short-term preference), so the importance of items for users may change. NARM and STAMP use attention mechanisms to adjust items' importance in sessions focusing on the last observed item and devising a new representation vector indicating the user's short-term preference. Finally, their model recommends items by combining the session representation vector (long-term preference vector) and the short-term preference vector. The difference between these two models is in considering the importance of vectors representing short-term and long-term preferences. In NARM, short-term and long-term preference vectors are concatenated, and the item scores are calculated by vectors passing through a linear layer. While in STAMP, the trilinear product of short-term and long-term vectors is used. While The principle of NARM and STAMP methods is to model the connection between the final item and the previous ones, the SR-GNN model \cite{wu2019session} is developed on the assumption that a significant part of information can be discovered by discovering more connections among all a session's items. Considering this idea, the SR-GNN uses a graphical representation of the session and a GNN-based approach to analyze it. 

\textit{Time Aware approaches:} Although these developed SBR deep models have achieved remarkable results, they all focus on extracting patterns from the sessions by scrutinizing the events observed during sessions. One of the first attempts to incorporate time information in SBR models is STAN \cite{garg2019sequence}. It examined the effect of three characteristics: the time interval between the current session and neighborhood sessions, the position of the item in neighborhood sessions, and the position of the item in the current session. K nearest neighbors of a session are determined using the cosine similarity between the item occurrence vectors of the session and the past ones. 

Some models have applied auxiliary information, including time intervals, to improve the performance of sequential recommendations. Multivariate Hawkes Process Embedding with attention (MHPE-a) \cite{wang2021sequential} maps each item and user entity to an embedding space. Then it obtains the probability that item $i$ will occur by user $u$ at time $t$ using a Multivariate Hawkes Process model and an attention mechanism. TiSASRec \cite{li2020time} employs absolute positions of items and time intervals between them in a user's sequence. For each user, a personalized time interval matrix is made, in which every matrix entity indicates a scaled time interval between two items. Then this matrix is used to adjust user and item embeddings. TIEN \cite{li2020deep} is a Click-Through Rate (CTR) prediction model introducing time-aware item behaviors in addition to user behaviors. The item behavior for an item is a sequence of users who interact with this item in chronological order. These item behaviors are adjusted by calculating the weight of each user using time intervals. However, since these models are designed for situations where user identifications are available, and user embedding is one of their main components, they cannot be directly applied for session based recommendation.

\section{Proposed Method: STAR}\label{sec:proposedModel}

In the STAR model, the learning process is performed in two main steps. First, we use the GloVe method \cite{pennington2014glove} to obtain initial item embeddings based on the co-occurrences of items in sub-session groups. Second, we fine-tune item embeddings, adjust them using time intervals between items, generate session representations, and predict the following items.

\begin{figure*}[t]
    \begin{center}
        \includegraphics[width=1\textwidth]{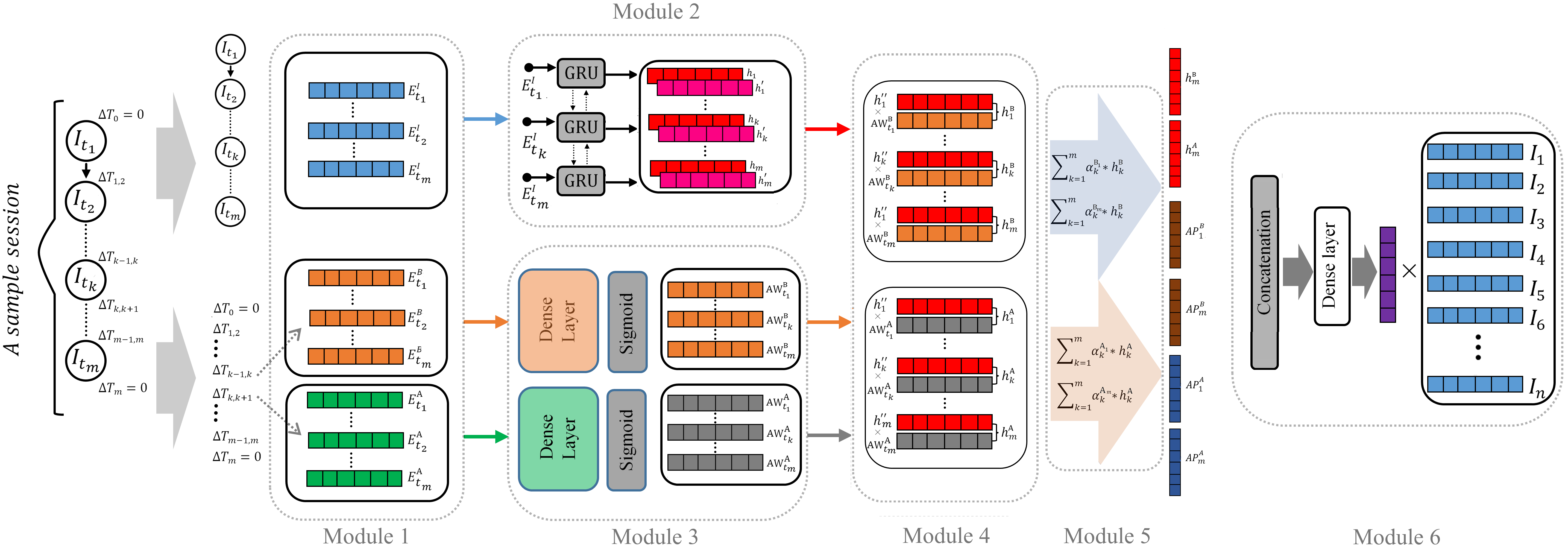}
        \caption{STAR Framework: A recommendation is generated by six modules in the STAR Framework. Module 1: Feature Extraction, Module 2: Session Encoding, Module 3: Weight Calculation,
        Module 4: Time Attention, Module 5: Self Attention, Module 6: Preference Detection.}
        \label{fig:STAR_framwork}
    
    \end{center}
\end{figure*}

\subsection{Problem Definition and Notation}
Suppose a typical user interacts with a set of items from steps $0$ to $m$, one item at each step. An SBR aims to predict the following item which this user will probably interact with at step $m+1>m$ based on her/his history of interactions (viewing items) from steps $0$ to $m$. Here we give a formulation of this problem:

Let $I=\{I_{1},I_{2},\dots ,I_{n}\}$ be the set of all items. Session $S=<I_{t_1},I_{t_2},\dots ,I_{t_m}>$, which $1<t_k<t_{k+1}<T$, is a sequence of items in chronological order from $I$ that a user has interacted with. Referring to (\ref{eq:score}), given a user interaction history in session $S$, our model (M) aims to predict the interaction probability of all $I_j \in I$  as the next preference of the user.

\begin{equation}
M(S) \approx {P(I_j | S) \quad \textrm{for every} \quad I_j \in I }
\label{eq:score}
\end{equation}

\begin{table}[!h]
\caption{Notation}
\label{tab:notation}
\centering
\begin{tabular}{ll}
\hline
Notation             & Description                                  \\ \hline
$I$                  & Item set                                     \\
$S$                  & A sample session of an anonymous user        \\
n                    & Number of items in $I$                       \\
d                    & Latent vector dimension                      \\
$X$                  & One-hot vector                               \\
$E$                  & Embedding vector.                            \\
$\Delta T_{k,{k+1}}$ & The time interval between value\$            \\
$h$, $h^{'}$         & Forward and backward session representations \\
$h^{''}$         & Aggregated session representations \\
AW                   & Attention weight vector                      \\
AP                   & Activated Preference vector                  \\
W                    & Weight matrix                                \\
b                    & Bias vector                                  \\
A                    & Stands for After                             \\
B                    & Stands for Before                            \\
hh                   & Stands for hour                              \\
mm                   & Stands for minute                            \\
ss                   & Stands for second
\\ \hline
\end{tabular}
\end{table}

\subsection{Extraction Initial Item Embeddings from Sub-Session Groups of Items}\label{sec:initial_embeddings}

\begin{figure}[h]
    \begin{center}
        \includegraphics[width=0.6\textwidth]{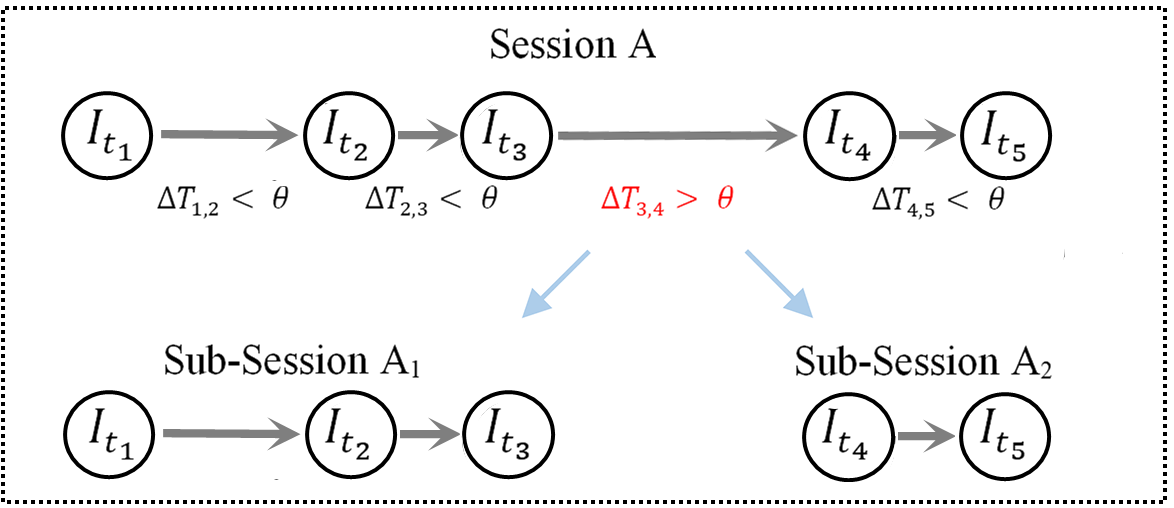}
        \caption{Breaking down the sessions into the sub-sessions}
        \label{fig:Session_Break}
    \end{center}
\end{figure}

As mentioned earlier, we use the GloVe technique to learn initial item embeddings. We consider this assumption that there may be some discontinuities inside a session, which means the session may comprise two or more conceptually independent sub-sessions, whereas the items inside a session are connected from the viewpoint of the active user. To break down each session into sub-sessions, we use the time intervals between user-item interactions in sessions. As illustrated in Fig.~\ref{fig:Session_Break} if the time interval between two items is longer than a threshold value $\theta$, this point is considered a breaking point. Next, sessions are broken (if possible), the GloVe co-occurrence matrix is constructed, and the initial representations of the items are obtained. Then, these representations are used as the initial embeddings in the next step of our approach.

\subsection{Learning Behavioral Patterns}
Learning behavioral patterns is the second step of the proposed algorithm. This part of the model consists of six modules, which are constructed in an end-to-end training fashion. These six modules are : 1) Feature Extraction 2) Session Encoding 3) Weight Calculation 4) Time Attention 5) Self Attention 6) Preference Detection (Fig.~\ref{fig:STAR_framwork}). In the following, we will explain each of these modules in more detail.

\subsubsection{Module 1, Feature Extraction}
For predicting the next item, the system receives the sequence of items visited so far in the session, as well as the time intervals between them. In the first step, it generates three kinds of outputs: A fine-tuned representation for each item and the vector representations for the time intervals observed before and after these items.  
As we will see later, the vector representations of time intervals will be combined with those of items to form the representation of the ongoing session. One can argue that each time interval appearing in a session can directly affect the role of the items before and after it, possibly in different ways. Similarly, an item's role in a session can be affected by the time intervals before and after it in different ways. So we use two separate embedding layers for time intervals to allow the system to interpret each time interval in these two different ways. 

To clarify how these components work, we will explain how this step's output is generated for the item $I_{t_k}$ in the sample session in Fig.~\ref{fig:STAR_framwork}. First, the one-hot vector ($X^{I}_{t_{k}}$) is considered for item $I_{t_k}$. Then, according to (\ref{eq:step1_0}), the item embedding $E^{I}_{t_{k}}$ is obtained by multiplying $X^{I}_{t_{k}}$ by the weights of the embedding layer $\mathcal{W}^{I} \in \mathbb{R}^{n*d}$, which was explained in section (\ref{sec:initial_embeddings}). 

\begin{equation}
	\begin{array}{l}
        E^{I}_{t_{k}} = X^{I}_{t_{k}} * \mathcal{W}^{I}
	\end{array}
\label{eq:step1_0}
\end{equation}

In the following, We calculate two time interval embeddings for this item, the time interval embeddings before and after the item, based on the values $\Delta T_{k-1,k}$ and $\Delta T_{k,k+1}$. First, we obtain unit 
embeddings. To do this, ($\Delta T_{k-1,k}$) and ($\Delta T_{k,k+1}$) are converted to "${hh}$:${mm}$:${ss}$" format. Next, one-hot vectors $\{X^{B_h}_{t_k},X^{B_m}_{t_k},X^{B_s}_{t_k}\}$ and  $\{X^{A_h}_{t_k},X^{A_m}_{t_k},X^{A_s}_{t_k}\}$ are considered for each value of "${hh}^B_{t_k}$:${mm}^B_{t_k}$:${ss}^B_{t_k}$" and  "${hh}^A_{t_k}$:${mm}^A_{t_k}$:${ss}^A_{t_k}$". Then, refer to (\ref{eq:step1_1_0}) and (\ref{eq:step1_1_1}), by multiplying these one-hot vectors by the embedding layer weights $\{{W}^{B_h} \in \mathbb{R}^{24*d},{W}^{B_m} \in \mathbb{R}^{60*d}, {W}^{B_s} \in \mathbb{R}^{60*d}\}$ and $\{{W}^{A_h} \in \mathbb{R}^{24*d},{W}^{A_m} \in \mathbb{R}^{60*d}, {W}^{A_s} \in \mathbb{R}^{60*d}\}$ unit embeddings are computed. 

Finally, refer to (\ref{eq:step1_2}), the concatenation of these embeddings is considered as the after and before time interval embeddings for item $I_{t_k}$. Because the number of time interval embeddings at step $k$ should be the same as the number of items in the session, we consider the time interval embeddings before the first item $E^{B}_{t_1}$ and after the last one $E^{A}_{t_k}$, but they are always zero vectors.
 
\begin{figure}[h]
\centering
\includegraphics[width=0.6\textwidth]{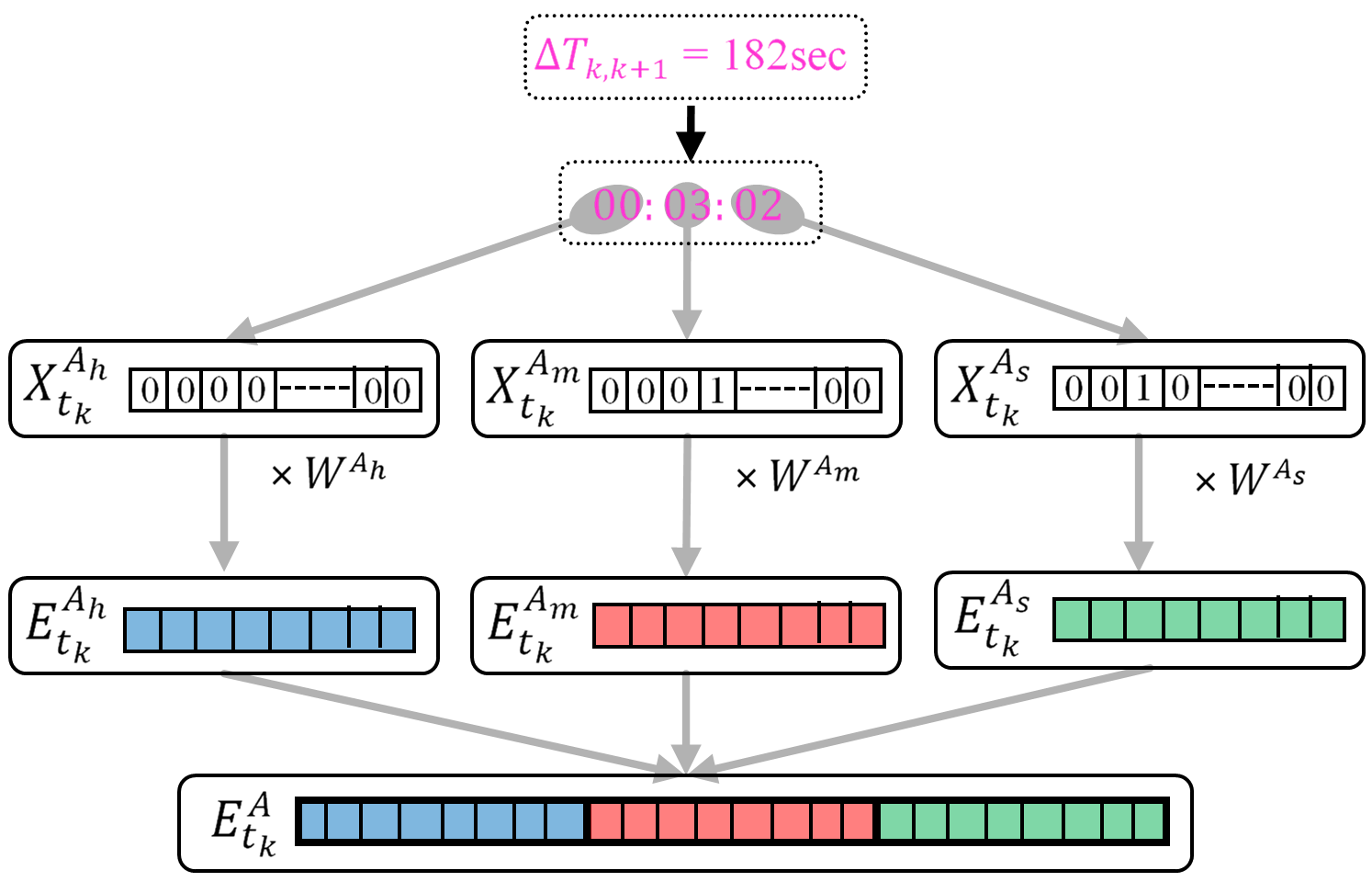}
\caption{Obtaining the after time interval embedding $E^{A}_{t_k}$ from the time interval value $\Delta T_{k,k+1}$.}
\label{fig:Time_Embedding}
\end{figure}

 Figure \ref{fig:Time_Embedding} depicts the process of converting the time interval value ($\Delta T_{k,k+1}=182$ seconds) to the time interval embedding after item $I_{t_k}$.

\begin{equation}
	\begin{array}{l}
        E^{B_{h}}_{t_k} = X^{B_h}_{t_k} * \mathcal{W}^{B_h} \\
        E^{B_{m}}_{t_k} = X^{B_m}_{t_k} * \mathcal{W}^{B_m} \\ 
        E^{B_{s}}_{t_k} = X^{B_s}_{t_k} * \mathcal{W}^{B_s} 
	\end{array}
\label{eq:step1_1_0}
\end{equation}

\begin{equation}
	\begin{array}{l}
        E^{A_{h}}_{t_k} = X^{A_h}_{t_k} * \mathcal{W}^{A_h} \\
        E^{A_{m}}_{t_k} = X^{A_m}_{t_k} * \mathcal{W}^{A_m} \\
        E^{A_{s}}_{t_k} = X^{A_s}_{t_k} * \mathcal{W}^{A_s} 
	\end{array}
\label{eq:step1_1_1}
\end{equation}

\begin{equation}
	\begin{array}{l}
	    E^{B}_{t_k} = [E^{B_{h}}_{t_k};E^{B_{m}}_{t_k};E^{B_{s}}_{t_k}] \\  
        E^{A}_{t_k} = [E^{A_{h}}_{t_k};E^{A_{m}}_{t_k}; E^{A_{s}}_{t_k}] 
    \end{array}
\label{eq:step1_2}
\end{equation}

\subsubsection{Module 2, Session Encoding}
This module encodes an aggregation of the events has happened so far in a session into a representation vector. The architecture of this module comprises one Bidirectional GRU \cite{schuster1997bidirectional}, whose inputs are item embeddings generated in the Feature Extraction module. The outputs are known as the session representations.

If $E^{I}_{t_k}$ is the embedding of item $I_{t_k}$ and $h_{k-1}$ and $h^{'}_{k-1}$ are the forward and backward session representations at step $t_{k-1}$, then according to (\ref{eq:step_2_0}), $h_{k}$ and $h^{'}_{k}$ will be forward and backward session representations up to step $t_k$. We consider the average of $h^{'}_{k}$ and $h_{k}$ as adjusted session representation at step $t_k$ (\ref{eq:step_2_1}).

\begin{equation}
	\begin{array}{l}
		z_k = \sigma(W_z \cdot [h_{k-1},E^{I}_{t_{k}}]) \\
		r_k = \sigma(W_r \cdot [h_{k-1},E^{I}_{t_{k}}]) \\
		\tilde h_k = tanh(W \cdot [r_k \ast h_{k-1}, E^{I}_{t_{k}}]) \\
		h_k = (1-z_k) \ast h_{k-1} + z_k \ast \tilde{h}_k \\ \\
		z^{'}_{k} = \sigma(W_z \cdot [h^{'}_{k-1},E^{I}_{t_{k}}]) \\
		r^{'}_{k} = \sigma(W_r \cdot [h^{'}_{k-1},E^{I}_{t_{k}}]) \\
		\tilde h^{'}_{k} = tanh(W \cdot [r^{'}_{k} \ast h^{'}_{k-1}, E^{I}_{t_{k}}]) \\
		h^{'}_{k} = (1-z^{'}_{k}) \ast h^{'}_{k-1} + z^{'}_{k} \ast \tilde{h}^{'}_k
	\end{array}
\label{eq:step_2_0}
\end{equation}  
 
\begin{equation}
	\begin{array}{l}
        h^{''}_{k} = (h_{k} + h^{'}_{k}) / 2
	\end{array}
\label{eq:step_2_1}
\end{equation}

\subsubsection{Module 3, Attention Weight Calculation}
In the Weight Calculation module, we calculate two sets of weights for session representations. $AW^{A}_{t_k}$ are $AW^{B}_{t_k}$  are attention weights obtained from after time interval embedding $E^{A}_{t_k}$ and before time interval embedding $E^{B}_{t_k}$. These weights are then utilized in the Time Attention module to change the importance of session representations at each step.

According to (\ref{eq:step_3}), these weights are obtained by passing the time interval embeddings through a linear layer ($W^{1}, W^{2} \in \mathbb{R}^{3d*d}$ $b^{1}, b^{2} \in \mathbb{R}^{d}$) and applying the Sigmoid ($\sigma$) activation function on them.

\begin{equation}
	\begin{array}{l}
        AW^{B}_{t_k} = \sigma((E^{B}_{t_k} * W^{1}) + b^1) \\
        AW^{A}_{t_k} = \sigma((E^{A}_{t_k} * W^{2}) + b^2)
     \end{array}
\label{eq:step_3}
\end{equation}

\subsubsection{Module 4, Time Attention}

This module is designed to adjust the user's behaviors in every step based on time interval weights. In this component, the weight vectors $AW^{A}_{t_k}$ and $AW^{B}_{t_k}$ are multiplied by the session representation vector $h^{''}_{k}$ in an elementwise way. This multiplication causes the amount of information in the $k^{th}$ step to be adjusted based upon the time intervals. The purpose of changing this step's information is to consider the assumptions mentioned in section (\ref{sec:Introduction}). Therefore, refer to (\ref{eq:step_4}), the output of this component will be the modified session representation vectors $h^{A}_{k}$ and $h^{B}_{k}$ per step. The symbol $\odot$ indicates an elementwise multiplication.

\begin{equation}
	\begin{array}{l}
        h^{A}_{k} = AW^{A}_{t_k} \odot h^{''}_{k} \quad h^{B}_{k} = AW^{B}_{t_k} \odot h^{''}_{k}
     \end{array}
\label{eq:step_4}
\end{equation}

\subsubsection{Module 5, Self Attention}
This module aims to identify the user's goal based on the first and the last items in a session by using the self-attention mechanism. This module's inputs are adjusted session representations $h^{A}_{k}$, $h^{B}_{k}$. The outputs of this component are the Activated Preference (AP) vectors. Based on (\ref{eq:step_5_1}), these vectors are obtained from the weighted combination of adjusted session representations. Our main difference from other research on the attention mechanism is that we apply the attention mechanism to adjusted session representation vectors.

\begin{equation}
	\begin{array}{l}
        AP^{A}_{1} = \sum_{k=1}^m \alpha^{A_1}_k * h^{A}_{k} \quad
        AP^{A}_{m} = \sum_{k=1}^m \alpha^{A_m}_k * h^{A}_{k}
        \\ \\
        AP^{B}_{1} = \sum_{k=1}^m \alpha^{B_1}_k * h^{B}_{k} \quad
        AP^{B}_{m} = \sum_{k=1}^m \alpha^{B_m}_k * h^{B}_{k} \\
     \end{array}
\label{eq:step_5_1}
\end{equation}

According to (\ref{eq:step_5_2}) weights of adjusted session representations are depends on the similarity between $h^{A}_{k}$ and $h^{A}_{1}$, $h^{A}_{m}$ and correspondingly, similarity between $h^{B}_{k}$ and $h^{B}_{1}$, $h^{B}_{m}$. We employ the softmax and dot product to calculate $\{ \alpha^{A_1}_k, \alpha^{A_m}_k, \alpha^{B_1}_k, \alpha^{B_m}_k \}$ as adjusted session representations' weights.

\begin{equation}
	\begin{array}{l}
        \alpha^{A_1}_k = softmax(h^{A}_{k} \cdot h^{A}_{1}) \quad \alpha^{A_m}_k = softmax(h^{A}_{k} \cdot h^{A}_{m})
        \\  \\
        \alpha^{B_1}_k = softmax(h^{B}_{k} \cdot h^{B}_{1}), \quad \alpha^{B_m}_k = softmax(h^{B}_{k} \cdot h^{B}_{m})
     \end{array}
\label{eq:step_5_2}
\end{equation}

\subsubsection{Module 6, Preference Prediction}

Given generated final session representations $\{h^{A}_{m}$, $h^{B}_{m}\}$ and the Activated Preference vectors $\{AP^{A}_{1}, AP^{A}_{m}, AP^{B}_{1}, AP^{B}_{m}\}$, this module aims to generate the item scores. Based on (\ref{eq:step6_1}), vectors $h$s and $AP$s are concatenated together. Then they pass through a linear layer with weights $\mathcal{W}^3 \in \mathbb{R}^{6d*d}$ and bias $b^{3} \in \mathbb{R}^{d}$.

\begin{equation}
    \begin{array}{l}
	z = [h^{A}_{m};h^{B}_{m};AP^{A}_{1};AP^{A}_{m};AP^{B}_{1};AP^{B}_{m}] \\
	z^{'} = (z * \mathcal{W}^3) + b^{3}
	\end{array}
\label{eq:step6_1}
\end{equation}

To generate item scores as a new user's preference, the linear layer output vector  $z^{'}$ is multiplied by the embedding layer weights $\mathcal{W}^{I}$ and added to bias $b^{4} \in \mathbb{R}^{n}$, then passed through a Softmax function.

\begin{equation}
	\begin{array}{l}
	    \tilde{Y} = Softmax( (z^{'} * \mathcal{W}^{I}) + b^{4}) = [\tilde{y}_1 ,\tilde{y}_2, \dots ,\tilde{y}_n ]
	\end{array}
\label{eq:step5_2}
\end{equation} 

We use multi class cross-entropy loss according to (\ref{eq:loss}). 

\begin{equation}
	\mathcal{L}(\tilde{Y}) = -\sum_{i=1}^n{y_i\log(\tilde{y}_i) + (1 - y_i)\log(1 - \tilde{y}_i)} 
\label{eq:loss}
\end{equation}

\section{Evaluation and Analysis}\label{sec:results}
This section first describes the datasets, the preprocessing step, baseline models, and evaluation metrics. Then we report experimental results for the STAR framework and the baseline models under two Recall@20 and MRR@20 evaluation metrics.

\subsection{Dataset and Preprocessing}
We evaluated the proposed framework on two datasets: YooChoose \cite{ben2015recsys} and Diginetica. YooChoose dataset contains sessions of an online webshop. Each session is composed of a set of clicking items for an unknown user. Some sessions also contain the buy events in addition to clicks, but like other studies, they were eliminated. The data has been collected during six months from an online retailer in Europe. We use the final day sessions for test data while the rest of the sessions are used for training. Because the Yoochoose dataset is quite large, like some previous studies (\cite{li2017neural, liu2018stamp, wu2019session}), we use $\frac{1}{4}$ and  $\frac{1}{64}$ of the training data.  

The Diginetica dataset is from CIKM Cup 2016. This dataset contains anonymized search and browsing logs, product data, anonymized transactions, and a large data set of product images. We use only transactional events for train and test. The sessions of the final week are used as the test set for Diginetica. 

In the preprocessing step, items appearing less than five times in the datasets and sessions with a length of less than two are removed. Also, uncommon items between test and train sets are excluded from the test set. Statistical information of both datasets is reported in Table ~\ref{tab:data_stat}.

\begin{table}[!h]
\caption{Statistics of Datasets Used in the Experiments}
\label{tab:data_stat}
\centering
\resizebox{.7\textwidth}{!}{
\begin{tabular}{@{}ccccc@{}}
\toprule
Dataset                                                                 & \# of clicks & \# of sequences\footnotemark & \# of items & Average length \\ 
\midrule
\textit{\begin{tabular}[c]{@{}c@{}}Yoochoose 1/64\\ Train\end{tabular}} & 486,614      & 369,859        & 17,370      & 6.39           \\ 
\midrule
\textit{\begin{tabular}[c]{@{}c@{}}Yoochoose 1/4\\ Train\end{tabular}}  & 7,841,282    & 5,917,745      & 30,445      & 6.02           \\ 
\midrule
\textit{\begin{tabular}[c]{@{}c@{}}Yoochoose\\ \textbf{Test}\end{tabular}}       & 71,222       & 55,898         & 6,751       & 6.72           \\ 
\midrule
\textit{\begin{tabular}[c]{@{}c@{}}Diginetica\\ Train\end{tabular}}     & 906,140      & 719,470         & 43,097      & 5.39           \\ 
\midrule
\textit{\begin{tabular}[c]{@{}c@{}}Diginetica\\ \textbf{Test}\end{tabular}}      & 76,821       & 60,858         & 21131       & 5.30           \\ 
\bottomrule
\end{tabular}}
\end{table}
\footnotetext[1]{Each session with length $m$ contains $m-1$ sequence(s).}

\subsection{Evaluation Metrics}
$Recall@k$ and $Mrr@k$ are evaluation metrics in this study.

\begin{enumerate}
	\item The $Recall@k$ specifies the average proportion of the target items that appear among the top-k recommended items. We have formulated this criterion as (\ref{Recall}). $TP@k$ is the number of times the target item is between top-K recommended items, and $FN@k$ is the number of times the target item is not between top-K recommended items.
	
	\begin{equation}
    	Recall@k = \frac{TP@k}{TP@k+FN@k}
    	\label{Recall}
    \end{equation}

	\item The $MRR@k$ measures the average reciprocal rank of the target item among top-k recommended items. If the target item does not appear in the top-k recommendation list, then the contribution of that recommendation to $MRR@k$ will be zero. In (\ref{MRR}), N is equal to the total test samples, and $rank_i$ is equal to the position of the target item in the top-k recommendation list.
	
	\begin{equation}
    	MRR@k = \frac{1}{N} \sum_{i=1}^N \frac{1}{\textrm{rank}_i} \quad \textrm{that} \quad 1\leq \textrm{rank}_i \leq k 
    	\label{MRR}
    \end{equation} 
	
\end{enumerate}

\subsection{Baseline Methods}
The result of this study is compared with the baseline methods described below.

\begin{itemize}

\item POP and SPOP: Pop recommends top-k most popular items. SPOP is similar to Pop but it places the items that occurred in the current session at the beginning of the top-k recommendation list based on the frequency of their occurrence in the current session.

\item Item-KNN \cite{sarwar2001item}: This baseline method recommends items similar to the items that the user has interacted with before in the current session. The cosine similarity measure is used to calculate item similarities.

\item BPR-MF \cite{rendle2012bpr}: This is a matrix factorization method. It optimizes a pairwise ranking objective function via SGD. To apply BPR-MF method in SBRs, the averages of the items that appeared in the session are calculated in the latent space, and the resulting vector is used as the user representation.

\item FPMC \cite{rendle2010factorizing}: This is an advanced hybrid model used for the next-basket recommendation. In order to use this model in SBRs, the user information is ignored.

\item GRU4Rec \cite{hidasi2016session}: This model that introduces the session parallel mini-batch method utilizes an RNN structure with pairwise ranking loss functions for the session-based recommendation.

\item NARM \cite{li2017neural}: Another RNN Based algorithm that also utilizes the attention mechanism to model the user's main purpose and behavioral pattern in the session. 

\item STAMP \cite{liu2018stamp}:
Captures long-term and short-term history of events in a session using an attention mechanism and MLP. Unlike NARM, STAMP explicitly emphasizes the current interest reflected by
the last click to capture the hybrid features of current and general interests \cite{liu2018stamp}.

\item SR-GNN \cite{wu2019session}: To make recommendations, this algorithm applies a GNN model over graph representations of sessions. It also uses an attention mechanism to combine the recent interests of the session with its global goals.

\item STAN \cite{garg2019sequence}: It examines the impact of the time interval between the current session and neighboring sessions, the position of the item in neighboring sessions, and the position of the item in the current session. K nearest neighbors of a session are determined using the cosine similarity between the item occurrence vectors of the session and the past ones.

\end{itemize}

\begin{table*}[t]
\centering
\caption{Performance Compared with Other Baselines}
\resizebox{.97\textwidth}{!}{
\begin{tabular}{ccccccccccccc}
\hline
\multirow{2}{*}{Methods} &  & \multicolumn{3}{c}{YOOCHOOSE 1/64}          &           & \multicolumn{3}{c}{YOOCHOOSE 1/4}           &           & \multicolumn{3}{c}{Diginetica}              \\ \cline{3-5} \cline{7-9} \cline{11-13} 
                         &  & Recall@20      &           & Mrr@20         &           & Recall@20      &           & MRR@20         &           & Recall@20      &           & MRR@20         \\ \hline
POP                      &  & 6.71           &           & 1.65           &           & 1.33           &           & 0.30           &           & 0.89           &           & 0.20           \\
S-POP                    &  & 30.44          &           & 18.35          &           & 27.08          &           & 17.75          &           & 21.06          &           & 13.68          \\
Item-KNN                 &  & 51.60          &           & 21.81          &           & 52.31          &           & 21.70          &           & 35.75          &           & 11.57          \\
BPR-MF                   &  & 30.30          &           & 12.08          &           & 3.40           &           & 1.57           &           & 5.24           &           & 1.98           \\
FPMC                     &  & 45.62          &           & 15.01          &           & -              &           & -              &           & 26.53          &           & 6.95           \\ \hline
GRU4REC                  &  & 60.64          &           & 22.89          &           & 59.53          &           & 22.60          &           & 29.45          &           & 8.33           \\
NARM                     &  & 68.32          &           & 28.63          &           & 69.73          &           & 29.23          &           & 49.70          &           & 16.17          \\
STAMP                    &  & 68.74          &           & 29.67          &           & 70.44          &           & 30.00          &           & 45.64          &           & 14.32          \\
SR-GNN                   &  & 70.57          &           & 30.94          &           & 71.36          &           & 31.89          &           & 50.73          &           & 17.59 \\
STAN                   &  & 69.45          &           & 28.74          &           & 70.07          &           & 28.89          &           & 50.97          &           & 18.48
\\ \hline
STAR                     &  & \textbf{71.31} & \textbf{} & \textbf{31.30} & \textbf{} & \textbf{72.46} & \textbf{} & \textbf{32.70} & \textbf{} & \textbf{53.98} & \textbf{} & \textbf{18.66} \\ \hline
\end{tabular}}
\label{tab:results}
\end{table*}

\subsection{Experiments}
We use validation data to fine-tune hyperparameters. Validation data is 10\% of the last sessions from train data. Hyperparameters setting are shown in Table ~\ref{tab:hyp_param}. The learning rate is chosen from $\{0.01, 0.001, 0.0001\}$. Suppose $l$ is the average time interval between sessions' items. We selected the most satisfactory $\theta$ value from $\{\frac{1}{3}l,\frac{1}{2}l,l,2l,3l\}$. We also utilized dropout layers for modules 1,2,3 and 6 after each weight layer to prevent overfitting. The best dropout rates, batch sizes and learning rate decay step for each dataset are reported in Table ~\ref{tab:Drops} which chosen from $\{0.1, 0.2, 0.3\}$ and $\{64, 128, 256, 512, 1024\}$ sets respectively. In all datasets, the value of $\theta$ is set twice the average time intervals in sessions. 

\begin{table}[h]
\caption{Parameters Setting}
\label{tab:hyp_param}
\centering
\begin{tabular}{@{}cc@{}}
\toprule
Parameter             & Value     \\ 
\midrule
Glove windows size & max session size         \\
First Step Training Epochs  & 100 epoch \\
Embedding Dimensions  & 180       \\ 
\midrule
Bidirectional GRU layers             & 1 layer   \\
Optimizer             & adam   \\
Learning rate         & 0.001      \\
Learning rate decay         & 0.1      \\
Total Epochs          & 12        \\
L2 penalty             & 1e-6 \\ \bottomrule
\end{tabular}
\end{table}

\begin{table}[!h]
\caption{Dropout Rates, Batch Size, and Learning Decay Step for Each Dataset.}
\label{tab:Drops}
\centering
\resizebox{.8\textwidth}{!}{
\begin{tabular}{lccc}
\hline
Dataset & Batch Size & Dropout & Learning rate Decay Step \\ \hline
Yoochoose 1/64 & 64         & 0.2     & 4\\ \hline
Yoochoose 1/4  & 512        & 0.1     & 4\\ \hline
Diginetica     & 512        & 0.3     & 6\\ \hline
\end{tabular}}
\end{table}

\subsection{Implementation Details}
STAR is implemented with \textit{PyTorch} and are available at the Github\footnotemark repository. All experiments are executed with a single GTX-1060 Ti GPU.

\footnotetext[1]{\href{https://github.com/yeganegi-reza/STAR}{https://github.com/yeganegi-reza/STAR}}

\subsection{Results}
The experiment results of the proposed model and baseline models are summarized in Table ~\ref{tab:results}. The results are reported for two statistic performance measures: $Recall@20$ and $MRR@k20$. The best results are highlighted. The results show that the performance of the suggested model in terms of both measures has improved compared to the baseline models. This shows the effectiveness of the proposed hypotheses and methods. 

The presented observations indicate that traditional models' performance is less efficient than models based on deep neural networks in SBRs. Poor performance of the POP model is due to ignoring the session information. Although the SPOP model considers the items appearing in the session, it only uses the frequency of items, and clearly, such a naive approach can not lead to satisfactory performance in the session-based recommendation.

Among the conventional models, Item-KNN has achieved relatively better results. This model does not make decisions based only on the last item of the session and utilizes the information of all items within the session. That is essential to discover the goal of the user.

Another interesting observation is that the Item-KNN also outperforms the FPMC model, a Markov chain-based model that assumes that the occurrence of items in a session is independent of the history of the session. This assumption is somewhat counter-intuitive and unreliable, leading to poor results as reported in Table ~\ref{tab:results}. Since we regard each session as one user in FPMC, we do not have enough memory to initialize it in Yoochoose $\frac{1}{4}$ dataset.

Although Matrix Factorization models have satisfactory results in general collaborative filtering domains, they have not performed well in SBRs. The main reason for the poor results of the BPR-MF model is probably the lack of appropriate user representations, which is an essential entity in MF approaches. 

Results show that the models using deep learning techniques outperform conventional models in the SBRs. All these models try to analyze the history of actions in the session for estimating the user's current interests and goals. GRU4Rec models the user's sequential behavior using an RNN-based approach, and the flexibility of its deep structure helps it model the user's behavior more accurately than traditional baselines. NARM and STAMP apply attention mechanisms to combine the short-term preferences of the user with his/her long-term goals. This approach improves their performance compared to GRU4Rec, and the results reveal that discovering the momentary interests of the user during a session can help us make significantly better recommendations. SR-GNN that achieves the best performance among the baseline algorithms applies a sophisticated graph-based approach to study the behavior of the user during the session more accurately. Its approach allows it to consider an item in the context of all other items in the session that clearly leads to better embeddings and a better recommendation performance. 

As the results show, the proposed STAR model outperforms all baseline methods, including deep models, in all experimental setups. STAR applies the main ideas from its predecessors; however, What makes it different from other baselines, is that it considers the time intervals between the events in the session while analyzing those events. STAR uses the temporal proximity of events inside the session to enhance the quality of initial item representations. It also embeds the time intervals to capture valuable information about the importance of items in the current mindset of the user. According to the reported results, embedding this information into the vector representation of an ongoing session helps STAR make more accurate predictions about the next event in the session.

\begin{table}[!h]
\centering
\caption{Results of Eliminating Attention Modules from The STAR Framework.}
\resizebox{.98\textwidth}{!}{
\begin{tabular}{ccccccccccccc}
\hline
\multirow{2}{*}{Model} &  & \multicolumn{3}{c}{YOOCHOOSE 1/64} &  & \multicolumn{3}{c}{YOOCHOOSE 1/4} &  & \multicolumn{3}{c}{Diginetica} \\ \cline{3-5}
                       &  & Recall@20      &      & Mrr@20     &  & Recall@20     &      & MRR@20     &  & Recall@20    &     & MRR@20    \\ \hline
STAR\_V1              &  & 71.22          &      & 30.98      &  &72.02               &      &31.03            &  & 53.58        &     & 18.39     \\
STAR\_V2               &  & 71.05          &      & 31.04      &  &72.25               &      &31.27            &  & 53.70        &     & 18.41     \\
STAR                 &  & \textbf{71.35}          &      & \textbf{31.35}      &  & \textbf{72.46}         &      & \textbf{32.70}      &  & \textbf{53.98}        &     & \textbf{18.66}     \\ \hline
\end{tabular}}
\label{tab:effects}
\end{table}

\subsection{Influences of the modules}
To analyzes the influence of Self Attention and Time Attention modules to the final results, we have presented the performance of the following versions of the algorithm with that of the original STAR model in Table (\ref{tab:effects})
\begin{itemize}

\item \textbf{STAR\_V1} : STAR without Self Attention module: $\{AP^{A}_{1}, AP^{A}_{m}, AP^{B}_{1}, AP^{B}_{m}\}$ vectors are eliminated from the framework. 

\item \textbf{STAR\_V2} : STAR without Time Attention module: $AW^{A}_{t_k}$ and $AW^{B}_{t_k}$ weight vectors are excluded from the framework, and solely the Self Attention module is applied to unadjusted session representations.
\end{itemize}

It can be observed in Table (\ref{tab:effects}) that eliminating any of attention modules from STAR decreases the performance. The Self Attention has a more important role in the algorithm's accuracy as the MRR criterion is lower in STAR\_V1 than STAR\_V2. This clarifies that the model tends to learn more accurate behavioral patterns by considering the users' main purpose in sessions. The results also imply that the Self Attention and Time Attention Modules are complimentary. This observation can indicate that session representations without time interval adjustment are not sufficiently authentic since Self Attention is applied to less-tuned session representation vectors. Furthermore, employing Time Attention without considering Self Attention which is adopted to capture the user's primary goal, leads to sub-optimal predictions.   

\section{Conclusion}\label{sec:conclusion}
In this paper, we presented an approach for using temporal information available in the session data to enhance the recommendation quality. We used the time feature to identify the discontinuities in the course of events in sessions. We embedded items based on their co-occurrence in the resulting sub-session groups, which may reflect the users' momentary interests. We also introduced an approach to embedding time intervals between the sessions' events that are used to adjust the weights of the items in the aggregated vector representation of sessions. Our experimental results indicated that the temporal information of sessions' events has the potential to help us build models that efficiently reflect the users' interests during their sessions. In this study, we focused on analyzing the temporal information inside sessions; however, as a subject for future research, it may be helpful to consider exploring temporal relations among events happening in different sessions to embed items and sessions so that they reflect the effect of trends and popular items as well.

\bibliography{STAR}
\end{document}